\documentclass{iaus}
\usepackage{graphicx}
\usepackage{natbib}			

\title[Identification of Spitzer-IRS staring mode targets in the Magellanic Clouds] 
{Identification of Spitzer-IRS staring mode targets in the Magellanic Clouds}

\author[Paul M.~E. Ruffle, Paul M. Woods \& F. Kemper]   
{Paul M.~E. Ruffle$^1$, Paul M. Woods$^2$, F. Kemper$^3$}

\affiliation{$^1$Jodrell Bank Centre for Astrophysics, Alan Turing Building, School of Physics and Astronomy, The University of Manchester, Oxford Road, Manchester M13 9PL, UK \\ email: {\tt paul.ruffle@manchester.ac.uk} \\[\affilskip]
$^2$Department of Physics and Astronomy, University College London, \\ Gower Street, London WC1E 6BT, UK \\[\affilskip]
$^3$Academia Sinica, Institute of Astronomy and Astrophysics, Taipei 10617, Taiwan}

\pubyear{2011} 
\volume{284}  
\pagerange{1--12}
\jname{The Spectral Energy Distribution of Galaxies}
\editors{R.J. Tuffs \&  C.C.Popescu, eds.}
\begin{document}

\maketitle

\begin{abstract}
The SAGE-LMC, SAGE-SMC and HERITAGE surveys have mapped the Magellanic Clouds in the infrared using the Spitzer and Herschel Space Telescopes. Over 8.5 million point sources were detected and catalogued in the LMC alone. Staring mode observations using the InfraRed Spectrograph (IRS) on board Spitzer have been obtained for 1,000 positions in the LMC and $\sim$250 in the SMC. From the infrared spectroscopy we have identified the nature of the sources for which spectroscopy is available. These IRS staring mode targets represent an important contribution to the SED of these dwarf galaxies. Here we report on our latest results.
\end{abstract}

\noindent
The SAGE survey \citep{meixner0612} detected and catalogued 8.5 million point sources in the Large Magellanic Cloud (LMC) with Spitzer's IRAC and MIPS instruments. 2.5 million point sources have been detected and catalogued \citep{gordon1110} in the Small Magellanic Cloud (SMC). The Spitzer archive contains 1,000 IRS staring mode observations within the SAGE-LMC footprint (including 197 from the SAGE-Spec legacy program) and $\sim$250 in the SMC. These targets are the brightest infrared point sources in both clouds, and as such represent an important contribution to the SED of these dwarf galaxies. We are now extending the initial classification of 197 sources in the LMC \citep{woods1103} to $\sim$1,250 Spitzer-IRS staring mode targets in the Magellanic Clouds.

This spectral classification will allow us to verify the larger photometric classifications that have come out of recent studies of the Magellanic Clouds \citep[e.g.][]{boyer1110}. By relating the spectral identification of objects to regions in colour-magnitude space, a grand picture of the stellar populations of both clouds can be obtained. Also, our spectroscopic classifications can be used to test photometric classification methods, e.g. those by \citet{boyer1110}. Moreover, since spectral features provide information on dust composition, we can investigate the physical and chemical environments in each cloud. Both the LMC and the SMC have sub-solar metallicities of 0.5 Z$_\odot$ and 0.2 Z$_\odot$ respectively \citep{boyer1110}. This affects the chemistry in circumstellar environments, both for molecular and dust species, thus altering the spectral appearance. Our classification of the IRS staring mode targets in the SMC will enable us to compare the chemistry to similar environments in the LMC, and get a clearer picture of dust formation in a metal-poor star-forming galaxy similar to those in the early Universe.

\begin{figure}
\center
\includegraphics[width=129mm]{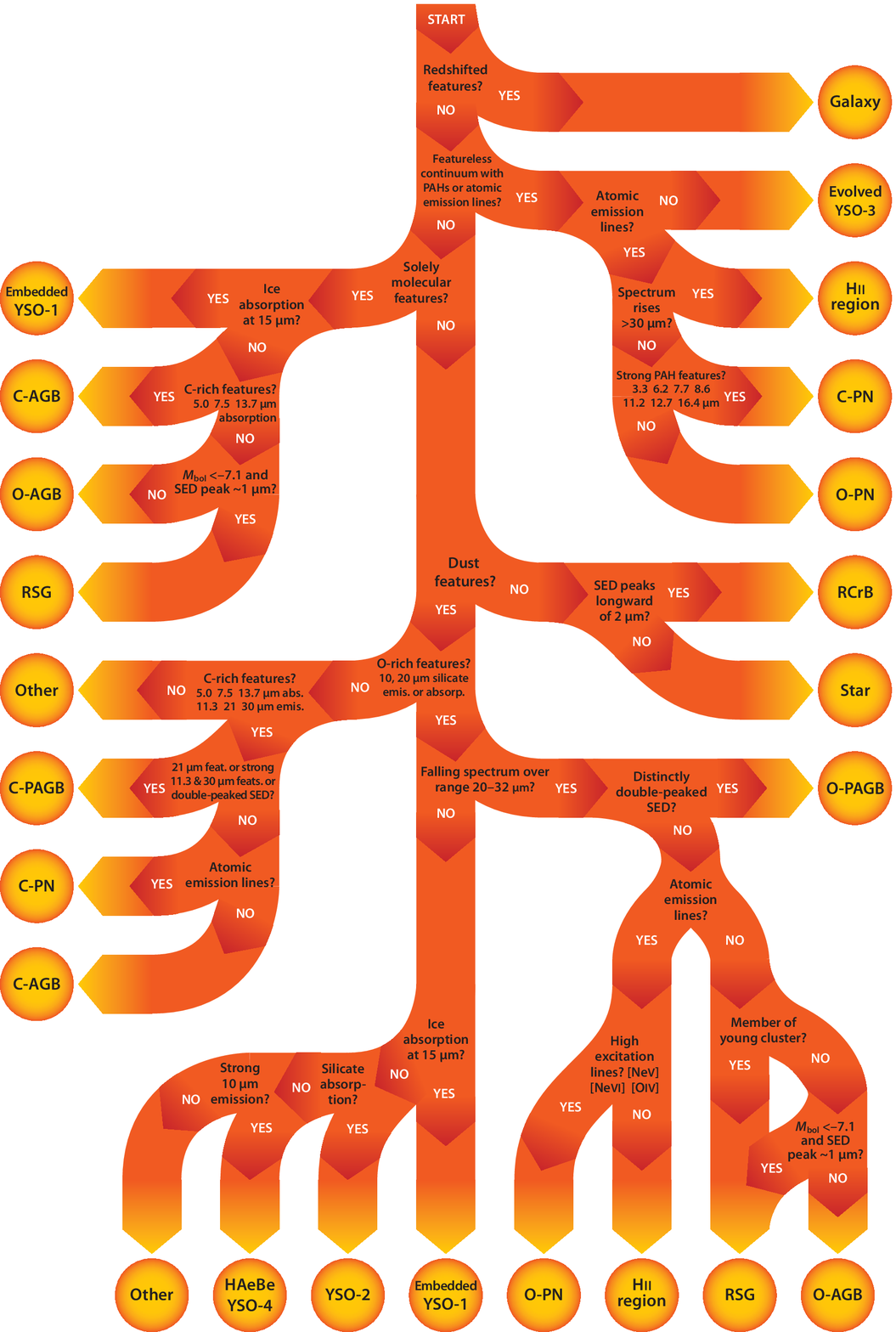} 
\caption{The logical steps of the redesigned classification decision tree of \citet{woods1103}, where 
         Spitzer IRS spectra ($\lambda$ = 5.2--38 $\mu$m), associated UBVIJHK, IRAC and MIPS photometry, 
         luminosity, variability, age and other information are used to classify LMC and SMC infrared 
         point sources. See Table~\ref{classgroups} for key to classification group codes.}
\label{dectree}
\end{figure}

Point sources are classified according to their Spitzer IRS spectra ($\lambda$ = 5.2--38 $\mu$m), associated UBVIJHK, IRAC and MIPS photometry, luminosity, variability, age and other information. Fig.~\ref{dectree} shows the redesigned classification decision tree of \citet{woods1103}, which we have used to classify each point source in our sample. For the LMC we find that we have 321 spectroscopically confirmed YSOs, 250 AGB stars, 105 {\sc Hii} regions, 68 post-AGB stars, 67 RSGs and 61 PNe. Objects are classified as per the groups in Table~\ref{classgroups} and a summary of our interim classifications for 1,000 LMC sources is listed in Table~\ref{classlmc} (Woods et al. \emph{in prep}). Work on the SMC is also in progress (Ruffle et al. \emph{in prep}).

\begin{table}
\caption{Classification groups used in the decision tree shown in Fig.~\ref{dectree}.}
\label{classgroups}
\setlength{\tabcolsep}{6.5pt} 
\begin{tabular}{ll|ll}
\hline 
Code	&  Object type				&  Code		&  Object type \\
\hline 
YSO-1	&  Embedded Young Stellar Objects	&  RCrB		&  R Coronae Borealis stars  \\
YSO-2	&  Young Stellar Objects		&  C-PAGB	&  Carbon-rich post-AGB stars  \\
YSO-3	&  Evolved Young Stellar Objects	&  O-PAGB	&  Oxygen-rich post-AGB stars  \\
YSO-4	&  HAeBe Young Stellar Objects		&  C-PN		&  Carbon-rich planetary nebulae  \\
Star	&  Stellar photospheres			&  O-PN		&  Oxygen-rich planetary nebulae  \\
C-AGB	&  Carbon-rich AGB stars		&  Other	&  Other known or unknown types  \\
O-AGB	&  Oxygen-rich AGB stars		&  {\sc Hii}	&  {\sc Hii} regions  \\
RSG	&  Red SuperGiants			&  Gal		&  Galaxies  \\
\hline 
\end{tabular}
\end{table}

\begin{table}
\caption{Summary of interim classifications for 1,000 LMC point sources.}
\label{classlmc}
\setlength{\tabcolsep}{7.25pt} 
\begin{tabular}{lr|lr|lr|lr}
\hline 
Type	& Count	&  Type			& Count	&  Type		& Count	&  Type		& Count \\
\hline 
YSO	&  321	&  RSG			&  67	&  C-PN		&  29	&  Unknown	&  8   \\
Star	&  35	&  C-PAGB		&  26	&  O-PN		&  32	&  Unclassified	&  78  \\
C-AGB	&  152	&  O-PAGB*		&  42	&  {\sc Hii}	&  105	&  		&      \\
O-AGB	&  98	&  (*inc. RVTau		&  9)	&  Galaxies	&  7	&		&      \\
\hline 
\end{tabular}
\end{table}

\begin{figure}[b]
\hbox{ 
\hspace{4mm}%
\includegraphics[width=56mm]{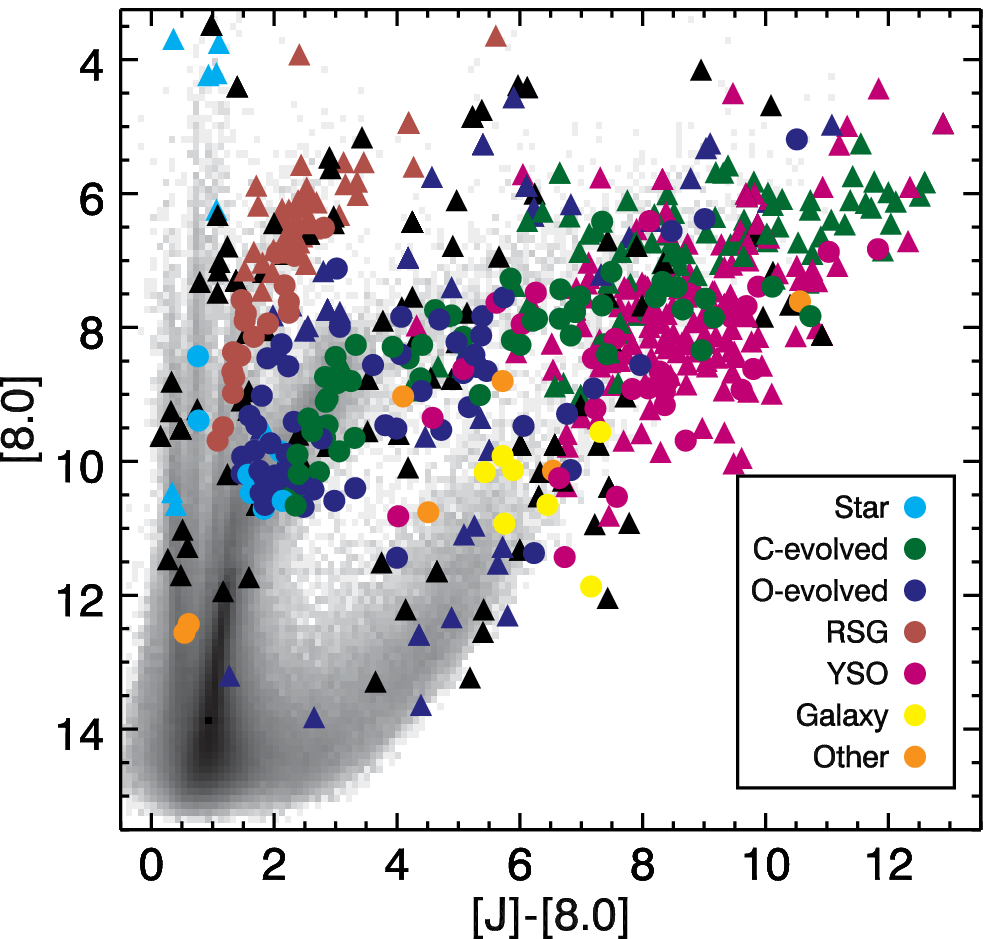}%
\hspace{5mm}%
\includegraphics[width=55.25mm]{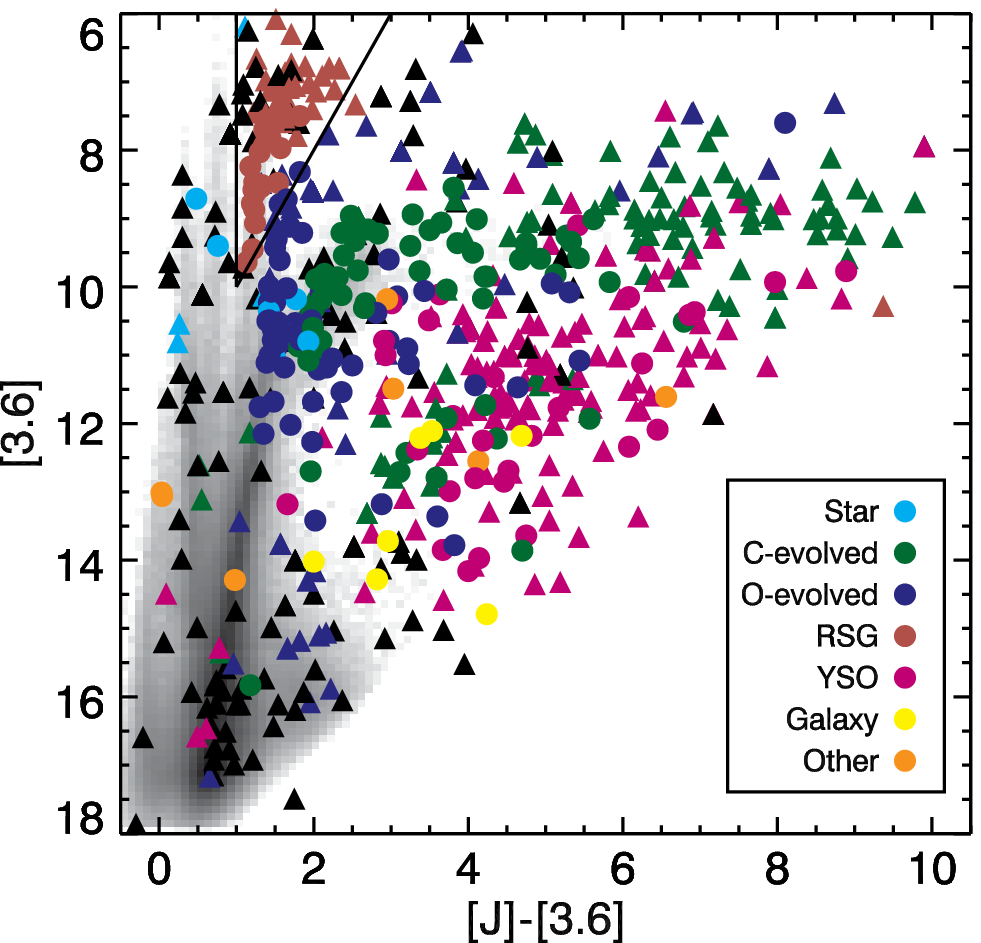}}
\caption{Example CMD distributions for 1,000 LMC point sources.
         Left: The initial 197 SAGE-Spec point sources from \citet{woods1103} are shown as filled 
         circles, with additional archival sources shown as filled triangles. 
         Black points are sources that still need to be classified.
         O-AGB, O-PAGB, O-PN, C-AGB, C-PAGB and C-PN classifications are merged into two broad groups.
         Right: The cut for RSGs can be clearly distinguished from other oxygen-rich evolved stars.
         The greyscale backgrounds show a Hess diagram of the entire SAGE-LMC sample.}
\label{cmds}
\end{figure}

\end{document}